\documentclass[prl,reprint]{revtex4-1}
\usepackage{bm}
\usepackage{graphicx}
\usepackage{epstopdf}
\begin{document}

\title{Local Plasticity as the Source of Creep and Slow Dynamics in Granular Materials}

\author{Ishan Srivastava}
\email[]{isrivast@purdue.edu}
\author{Timothy S. Fisher}
\email[]{tsfisher@purdue.edu}

\affiliation{School of Mechanical Engineering and Birck Nanotechnology Center, Purdue University, West Lafayette, IN 47907, USA}

\date{\today}

\begin{abstract}
Creep mechanisms in uniaxially compressed 3D granular solids comprised of faceted frictionless grains are studied numerically using a constant pressure and constant stress simulation method. Rapid uniaxial compression followed by slow dilation is predicted on the basis of a logarithmic creep phenomenon.  Micromechanical analysis indicates the existence of a correlation between granular creep and grain-scale deformations. Localized regions of large strain appear during creep and grow in magnitude and size with time. Furthermore, the accumulation of non-affine granular displacements increases linearly with local strain, thereby providing insights into the origins of plastic dissipation during stress-driven creep evolution. The prediction of slow logarithmic dynamics in the absence of friction indicates a universality in the role of plastic dissipation during the creep of granular solids.
\end{abstract}

\maketitle

Long-time evolution of granular materials under external stress is of significant importance in many natural and artificial processes. Time-dependent aging and creep in soils~\cite{Schmertmann1991} and concrete~\cite{Vandamme2009} under uniaxial stress are critically associated with the structural stability of man-made and natural geological constructs. The arrest (\textit{jamming}) of dry cohesionless granular materials upon densification~\cite{OHern2003,Majmudar2007} and/or the application of external stress~\cite{Bi2011,Dagois-Bohy2012} has been numerically and experimentally investigated. Concomitantly, external stress can also initiate the yielding (\textit{unjamming}) of granular packings \cite{DaCruz2005}. Critical state theories in soil mechanics \cite{Nedderman:1992aa} have proposed rate-independent, Mohr-Coulomb-type friction laws to describe the yielding of granular solids, and more recently, rate-dependent constitutive laws \cite{Jop2006} have improvised a description of dense granular flows above yield stress. However, a detailed description of the internal dynamics in granular materials below the yield stress is lacking \cite{Balmforth2014}. 

Diverse experiments at various length scales---from colloidal \cite{Coussot2002} to bio-cellular \cite{Bursac2005} to geological \cite{Houssais2015}---on kinematically constrained, glassy materials have demonstrated aging and creep relaxation at stresses below their yield strength.  Experiments on granular samples have demonstrated slow aging with time and creep when subjected to a constant external stress \cite{Nguyen2011,Amon2012} or a quasistatically increasing stress in stick-slip shearing experiments \cite{Hartley2003,Nasuno1998}. Logarithmically slow densification has also been observed in various experiments on compaction in confined granular systems \cite{Brujic2005,Richard2005,Knight1995}. 

Several explanations and models have been proposed for mechanically induced aging behavior in granular systems. Hartley and Behringer \cite{Hartley2003} suggested that collective irreversible granular rearrangements are responsible for slow dynamics. The emergence of such collective irreversibility in amorphous materials is often explained through long-ranged elastic coupling of localized plastic events \cite{Maloney2006,Schall2007}. Macroscopically, this complex spatio-temporal phenomenon is often condensed into a dynamical internal state variable called fluidity \cite{Kamrin2012}, based on the theoretical formulations of soft glassy rheology \cite{Bocquet2009}. On the other hand, other explanations attribute slowly evolving dynamics to the gradual strengthening (or aging) of the individual solid-friction contacts with a slow increase in the local coefficient of friction \cite{Kuhn1993}. 

In this letter, we simulate the temporal evolution of the micromechanics of granular packings comprised of faceted grains that are confined at a constant pressure and subjected to a constant uniaxial compressive stress below the yield limit. The simulations reveal logarithmically slow creep after initial exponentially rapid dynamics. The influence of structural transformations on creep rheology is assessed by decomposing the local grain-scale motions into elastic and plastic contributions. Initial exponential dynamics are governed by elastic deformations, and plastic transformations dominate granular motion as the system transitions into a slow creep. The accumulated local plasticity in stressed granular packings---characterized by non-affine granular motions---is observed to be strongly correlated with a slow creep at the macroscale. We hypothesize that the concert of collective granular dynamics arising from a complex interplay of local elastic and plastic deformations is the source of slow creep dynamics and it is a hallmark of the general class of amorphous solids.
 
\begin{figure}[t]
 \includegraphics{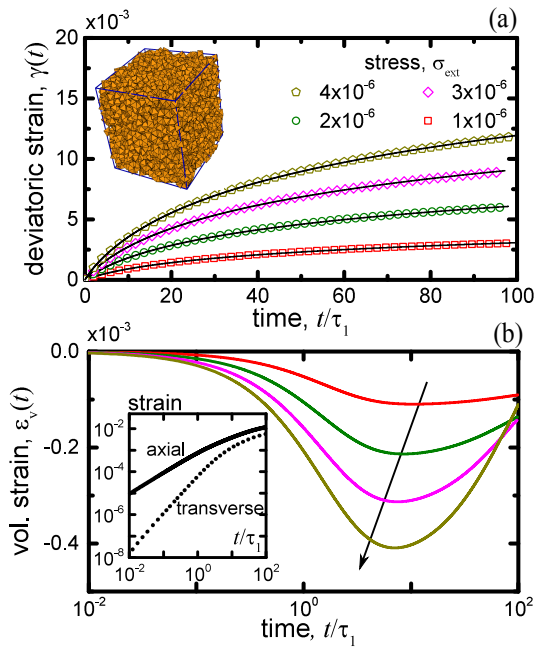}
 \caption{\label{global} (a)  Evolution of deviatoric strain $\gamma(t)$ for four different uniaxial compressive stresses. The solid black curves represent fits to Eq.~(\ref{eq3}) [inset: a selected realization of jammed octahedral grains. Dark lines indicate periodic-cell boundaries]. (b) Volumetric strain evolution $\epsilon_{v}(t)$ for four different uniaxial compressive stresses $\sigma_{\text{ext}}$ (see legend in (b) for color codes; arrow represents increasing stress) [inset: evolution of compressive strain $\epsilon_{xx}$ along the axial direction and tensile strain $\left(\epsilon_{yy}+\epsilon_{zz}\right)$ along the transverse direction of the periodic cell boundaries relative to the direction of applied external stress].}
 \end{figure}

\paragraph{Simulation Details.}Initially, elastic and frictionless grains were jammed under a hydrostatic pressure $P$ through a recently introduced enthalpy-based, variable-cell jamming simulation method~\cite{Smith2014}. The system is three-dimensional and contains $4096$ monodisperse, octahedral-shaped grains, as shown in the inset of Fig.~\ref{global}(a). A very small ratio of the hydrostatic pressure and the material elastic modulus of the grains is prescribed here to represent the limit of hard granular systems \cite{DaCruz2005}. Repulsive pair potentials describe the elastic force $\mathbf{F}^{\text{e}}$ and moment $\mathbf{M}^{\text{e}}$ between contacting grains (for details on force and moment calculations, see~\cite{Smith2010}). 

To investigate the creep phenomenon in jammed granular systems under a constant external uniaxial stress $\bm{\sigma}_{\text{ext}}$, we introduce dynamics into the motion of grains and the periodic cell. The magnitude of the {\it{xx}} component of the stress tensor is $\sigma_{\text{ext}}$ and the other components are zero. The geometry of the periodic cell is described by a metric tensor $\bm{g}=\bm{h}^{T}\bm{h}$, where $\bm{h}$ is a matrix whose columns are the Bravais lattice vectors of the periodic cell. Translational motion of a grain $i$ is described in lattice coordinates $\bm{s}_{i}=\bm{h}^{-1}\bm{r}_{i}$ instead of Cartesian coordinates $\bm{r}_{i}$, and the rotation $\bm{\omega}_{i}$ is described relative to its center of mass. Dissipation during granular motion is modeled using a mean-field dissipation model \cite{Vagberg2014} described in detail in the accompanying supplementary material \footnote{See Supplemental Material for detailed derivation and dimensional analysis of the equations of motion.}. 

Under low external stresses, the creep progresses at low strain rates, and the inertial effects of granular and periodic cell motion are assumed to be negligible in comparison to elastic interactions and dissipation. The equations of motion of grain displacement, grain rotation and periodic cell deformation are derived from a Lagrangian formulation \cite{Note1}. The equations of non-affine displacement and rotation of a grain $i$ are given as: 
\begin{equation}
D_{g}\dot{\bm{s}}_{i} = \widetilde{\mathbf{F}}_{i}^{e} \qquad\text{and}\qquad D_{g}\dot{\bm{\omega}}_{i} = \mathbf{M}_{i}^{e},
\label{eq1}
\end{equation}

where $\widetilde{\mathbf{F}}_{i}^{\text{e}}$ is the elastic force in lattice coordinates \cite{Souza1997}, $\mathbf{M}_{i}^{\text{e}}$ is the net moment on grain $i$ due to all the contacting grains, and $D_{g}$ is the dissipation constant for granular motion. The equation of motion of the periodic cell in the limit of negligible inertia is defined as: 
\begin{equation}
D_{c}\dot{\bm{g}} + \left(\widetilde{\bm{\sigma}}_{\text{app}} - \widetilde{\bm{\sigma}}_{\text{int}}\right) = 0.
\label{eq2}
\end{equation}
Here, $\widetilde{\bm{\sigma}}_{\text{int}}$ is the internal stress tensor \cite{Smith2014} in lattice coordinates and $\bm{\sigma}_{\text{app}}=\bm{\sigma}_{\text{ext}} + P\mathbf{I}$ is the total applied stress, where $\mathbf{I}$ is the identity matrix. The applied stress tensor in lattice coordinates is denoted by $\widetilde{\bm{\sigma}}_{\text{app}}$ \cite{Note1}. The dissipation constant for periodic cell motion is denoted by $D_{c}$. 

The equations of motion, Eqs.~(\ref{eq1}) and~(\ref{eq2}), were integrated using the Heuns second order predictor-corrector method. The units of time were chosen such that all dissipation constants are unity. All  displacements and volumes are scaled by grain size and volume respectively, and the stresses are scaled by the material elastic modulus of the grains. 

\paragraph{Macroscopic Evolution.} Uniaxial compressive stress was applied to the granular packing of $4096$ octahedral grains jammed at pressure $P=10^{-4}$ and the macroscopic strain tensor $\bm{\epsilon}$ was calculated from the deformation of the periodic cell boundaries \cite{Smith2014}. We restrict our analysis to the first and second invariants of the tensor: volumetric strain $\epsilon_{v}=\frac{1}{3}\sum_{k}\epsilon_{kk}$ and deviatoric (shear) strain $\gamma = \sqrt{\frac{3}{2}\sum_{i,j}\left(\epsilon_{ij}-\epsilon_{v}\delta_{ij}\right)}$, where $\delta_{ij}$ is the Kronecker delta tensor. The evolution of deviatoric strain with time is well described by two-stage dynamics represented phenomenologically by: 
\begin{equation}
\gamma(t)=\gamma_{0}\left(1-e^{-t/\tau_{1}}\right) + \gamma_{1}\text{log}\left(\frac{t}{\tau_{2}}+1\right),
\label{eq3}
\end{equation}
where $\tau_{1}$ is a viscoelastic time constant, and $\tau_{2}$ sets the timescale for slow dynamics. Fig.~\ref{global}(a) displays the evolution of deviatoric strain with time (normalized by the fast dynamics time constant $\tau_{1}$) at four different compressive stresses and the same jamming pressure. Black lines indicate best fits to Eq.~\ref{eq3}, and good agreement is observed at both long and short times. The first term of Eq.~\ref{eq3} denotes an initial Kelvin viscoelasticity in which $\gamma_{0}$ is the total elastic strain that is delayed by a viscous time constant $\tau_{1}$~\cite{Findley:1976aa}. The second term corresponds to a logarithmic creep that is caused by long-term structural rearrangements and this phenomenon has been previously observed in colloidal glasses~\cite{Siebenburger2012} and dense granular materials~\cite{Nguyen2011,Amon2012}. Such distinct relaxation regimes corresponding to fast and slow dynamics has been previously reported in experiments on granular compaction \cite{Knight1995,Brujic2005}.  

The granular packings exhibit rapid compaction at short times, whereas slow dilation occurs at longer times (see Fig.~\ref{global}(b)). Because the variable-cell method enables full tensorial evolution of strain while maintaining a constant external stress, the temporal response of transverse and axial strains at all times can be predicted (see inset of Fig.~\ref{global}(b)). At short times the transverse tensile strain $\left(\epsilon_{yy}+\epsilon_{zz}\right)$ is negligible, and rapid granular compaction is caused entirely by axial compression $\epsilon_{xx}$. During slow creep the tensile transverse strain competes with the compressive axial strain and results in slow dilation within the packing. This prediction correlates well with experimental observations during displacement pile setup in dense sands---comprised of angular grains---in which an initial volumetric contraction was followed by a long-term kinematically restrained dilation \cite{Bowman2005}.

 \begin{figure}[t!]
 \includegraphics{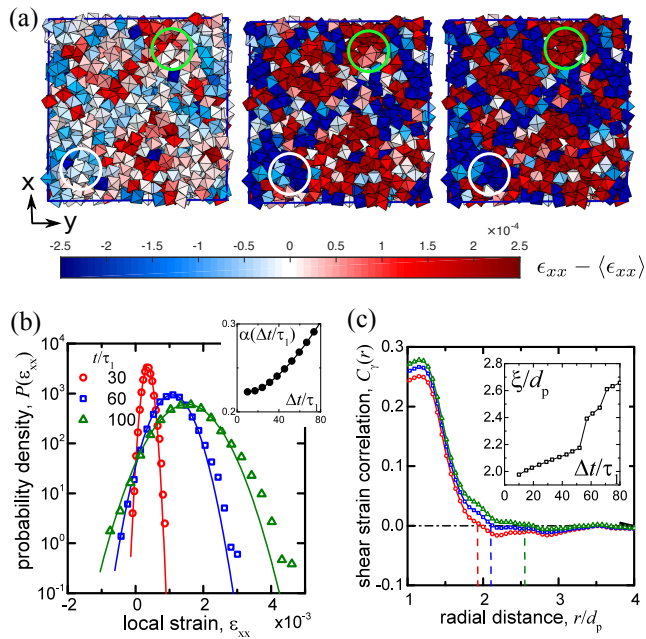}
 \caption{\label{local} (a) Deviations in the compressive strain $\epsilon_{xx}$ along a {\it{xy}}-plane at three times (left to right: $t=30\tau_{1}$, $60\tau_{1}$ and $100\tau_{1}$) relative to a reference configuration at $t=20\tau_{1}$ in the slow regime. White and green circles indicate growing regions of extensional and compressive strain respectively. (b) Probability distributions of local compressive strains $\epsilon_{xx}$.~The solid lines are Gaussian fits [inset: non-gaussian parameter $\alpha$ at various time intervals]. (c) Spatial correlation function of local deviatoric strain $C_{\gamma}(r)$ for three time intervals (see (a)) as a function of radial distance $r$ normalized by the in-sphere diameter $d_p$ of an octahedral grain [inset: normalized correlation length $\xi$ at various time intervals].}
 \end{figure}
 
\paragraph{Heterogeneous Micromechanics.}To understand the microscopic mechanisms that govern global creep response, the complex spatio-temporal granular dynamics were decomposed in terms of local structural rearrangements. A time-dependent local strain for grain $i$ was determined by the best-fit affine matrix $\mathbf{F}_{i}$ that transforms the center-to-center vectors $\mathbf{r}_{ij}$ of $n$ contacting grains over a time step of $\Delta t$ by minimizing $D_{i}^{2} = \sum_{j=1}^{n}\left[\mathbf{r}_{ij}\left(t+\Delta t\right) - \mathbf{F}\mathbf{r}_{ij}\left(t\right)\right]^2$ \cite{Falk1998} . A symmetric Lagrangian strain tensor can then be calculated as $\bm{\epsilon}_{i}=\frac{1}{2}\left(\mathbf{F}_{i}^{T}\mathbf{F}_{i}-\mathbf{I}\right)$. The remaining non-affine component of motion $D_{i}^{2}$ represents a component of the granular motion in the system that is elastically irreversible. 

In the regime of slow granular creep, a growing heterogeneity of the local strain field is observed. With a reference configuration at $t=20\tau_{1}$, we focus on the deviations of the uniaxial strain component  $\epsilon_{xx}$ (parallel to the global uniaxial strain) along a {\it{xy}}-plane of the packing and illustrate its distribution relative to the reference configuration for $t=30\tau_{1}$, $60\tau_{1}$ and $100\tau_{1}$ in Fig.~\ref{local}(a). The grains are color-coded based on the local uniaxial strain. White and green circles indicate regions of tensile and compressive uniaxial strain, respectively. We observe that the regions of negative and positive strain extend over many grains, and the spatial extent of these regions grows with time. Furthermore, the regions of positive (negative) strain are more positively (negatively) strained with the progress of time, as indicated within the white (green) circles. These regions indicate zones of large strain localization that are analogous to shear transformation zones in other amorphous materials \cite{Schall2007,Wang2013}. 

The distribution of local strains broadens with time, as shown in Fig.~\ref{local}(b). Furthermore, the strain distribution becomes less Gaussian with time, as indicated by the non-Gaussian parameter $\alpha(\Delta t)$ in the inset of Fig.~\ref{local}(b). This phenomenon is a signature of increasingly correlated transformations wherein localized deformations are coupled across multiple grains. To quantify this coupling, we compute the spatial autocorrelation of local deviatoric strain $\gamma$, defined as:
\begin{equation}
C_{\gamma}(\Delta r) = \frac{\langle \gamma(r+\Delta r)\gamma(r)\rangle -\langle\gamma(r)\rangle^{2}}{\langle \gamma(r)^{2}\rangle - \langle \gamma(r)\rangle^{2}},
\label{eq4}
\end{equation}
 where angular brackets denote ensemble averages. As shown in Fig.~\ref{local}(c), as time progresses in the slow creep regime, the local strains become more spatially correlated with an increasing correlation length as indicated in the figure inset. The correlation length is defined as the radial distance at which the autocorrelation goes to zero (denoted with dashed lines in the figure). This correlated deformation has been previously identified as a signature of the elastic response of glassy materials to local shear transformation zones \cite{Jensen2014} via stress redistributions caused by Eshelby fields \cite{Chattoraj2013}. 

  \begin{figure}[t]
 \includegraphics{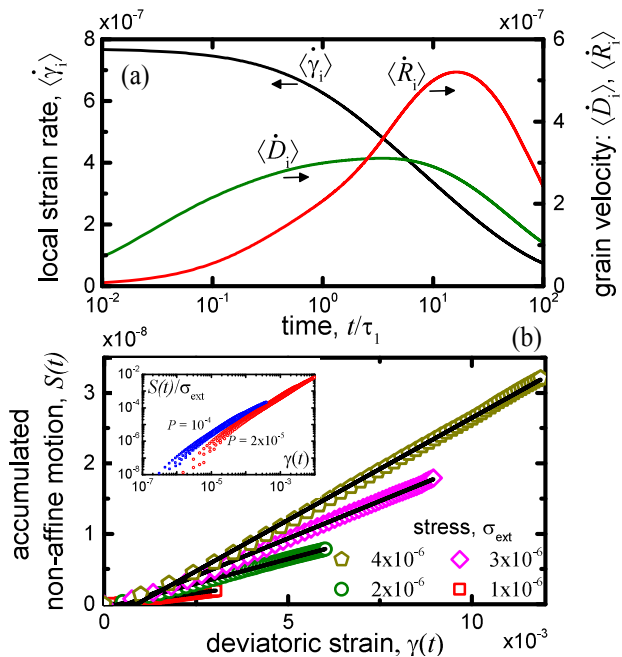}
 \caption{\label{motion} (a) Temporal evolution of the ensemble average of local deviatoric strain rate $\langle \dot{\gamma}_{i} \rangle$ (black), non-affine velocity $\langle \dot{D}_{i} \rangle$ (green) and rotational velocity $\langle \dot{R}_{i} \rangle$ (red) as a function of time. (b) Accumulated non-affine motion $S(t)$ as a function of global deviatoric strain $\gamma(t)$ for four uniaxial compressive stresses (see legend) and a jamming pressure $P=2\times 10^{-5}$. The solid black lines are linear fits. Inset: Scatter plots of the accumulated non-affine displacements divided by the applied uniaxial compressive stress for two different jamming pressures. Red circles represent data from the main graph. Blue rectangles represent data for $P=1 \times 10^{-4}$ and applied uniaxial stresses: $\sigma_{\text{ext}} = 1\times 10^{-7}$,  $2\times 10^{-7}$, $3\times 10^{-7}$ and $4\times 10^{-7}$.}
 \end{figure}
 
\paragraph{Evolution of Plastic Activity.}To elucidate the correlations between macroscale creep and local transformations, we quantify the temporal evolution of local dynamics at the grain scale by calculating the ensemble average of local deviatoric strain rate $\langle\dot{\gamma}_{i}\rangle$, non-affine velocity $\langle \dot{D}_{i} \rangle$ and rotational velocity $\langle \dot{R}_{i} \rangle$ as functions of time.  Granular rotation is quantified by the magnitude of the rotation of a grain about its instantaneous axis of rotation that is given by the cross product of unit vectors that connect the grain centroid to a specific vertex between consecutive time steps. 

Fig.~\ref{motion}(a) shows that at short times, when the macroscopic dynamics are exponential ($t \ll \tau_{1}$), local mechanical transformations are caused almost entirely by affine elastic deformations in the packing, as indicated by large shear strain rates and low non-affine and rotational velocities. This result indicates that the early transformations are essentially elastic in nature with little irreversibility in the granular motion. However, as the granular packing transitions into the logarithmic slow creep regime ($t \sim \tau_{1}$), it experiences increased irreversibility as evidenced by a decrease in local shear strain rates, and an increase in the non-affine and rotational velocities. During long periods of slow creep ($t \gg \tau_{1}$), both the elastic and non-affine contributions to local deformations decay slowly as the system creeps towards a metastable state. The close association between the growth of local non-affine transformations and the evolution of slow macroscopic creep indicates the presence of a causal relationship.

This relationship was further explored by correlating the temporal accumulation of squared non-affine displacements with the temporal evolution of macroscopic strain. Quantitatively, this accumulation at any time $t$ is computed as $S(t) =  S(t-\Delta t)+ \langle D_{i}^{2}\left(t-\Delta t,t\right) \rangle$, where the angular brackets denote the ensemble average of $D_{i}^{2}$ recorded between times $t-\Delta t$ and $t$.  Fig.~\ref{motion}(b) displays the relationship between cumulative non-affine displacements and macroscopic creep strain. The relationship is linear (except at short times), and the slope decreases with decreasing magnitude of applied uniaxial stress. Furthermore, all the curves collapse to a single linear curve when $S(t)$ is divided by the applied stress $\tau$ (see inset for two jamming pressures). This relationship can be expressed as $S(t) \sim \sigma_{\text{ext}} \gamma(t)$, and it reinforces the evidence that local irreversible deformations play a dominant role in the global creep process. We also note that for granular systems, jamming pressure plays a crucial role in plastic dissipation. From the inset of Fig.~\ref{motion}(b), the scale of plastic dissipation per unit elastic relaxation is larger in systems jammed under larger hydrostatic pressure. This result is expected because the large elastic modulus of highly jammed grains increases elastic coupling between plastic events.  

Lastly, we provide a mechanical definition for $S(t)$ by considering $\sigma_{\text{ext}} \gamma(t)$ to represent the elastic energy density released in the bulk during the creep relaxation process at long times. In athermal systems, elastic energy relaxation should be balanced by plastic dissipation in the system. Therefore, $S(t)$ represents the scale of plastic dissipation density in the system that is driven by external stress and manifests itself in the form of non-affine granular displacements. External stress-facilitated accumulation of non-affine displacements was also previously demonstrated for thermally driven amorphous metallic glass systems \cite{Wang2013}. 

\paragraph{Conclusion.} This letter identifies the internal structural transformations that govern the creep relaxation of uniaxially compressed granular solids. A full tensorial analysis of the system dynamics, enabled by a variable-cell simulation method, provides constitutive prediction of the temporal evolution of deviatoric and volumetric strains at all scales. The simulations predict rapid uniaxial compression at short times followed by slow dilation at long times. The accumulation of irreversible granular motion is proposed as a micromechanical explanation for slow logarithmic creep and is directly correlated with the plastic dissipation density within the system. Importantly, a simplified mean-field dissipation model captures the slow logarithmic creep dynamics observed in real granular solids \cite{Nguyen2011} and glassy colloids \cite{Siebenburger2012}, thereby indicating a universality in the role of localized plasticity during slow relaxation of amorphous solids. Future work will focus on the effects of more complex grain shapes and inter-granular friction on creep relaxation processes. 

\paragraph{Acknowledgements.} This work was supported by Grant No.~1344654 from the National Science Foundation Scalable Nanomanufacturing Program. We also acknowledge useful discussions with Prof M. Ashraf Alam. 

\bibliographystyle{apsrev4-1}
\bibliography{main}

\begin{thebibliography}{37}%
\makeatletter
\providecommand \@ifxundefined [1]{%
 \@ifx{#1\undefined}
}%
\providecommand \@ifnum [1]{%
 \ifnum #1\expandafter \@firstoftwo
 \else \expandafter \@secondoftwo
 \fi
}%
\providecommand \@ifx [1]{%
 \ifx #1\expandafter \@firstoftwo
 \else \expandafter \@secondoftwo
 \fi
}%
\providecommand \natexlab [1]{#1}%
\providecommand \enquote  [1]{``#1''}%
\providecommand \bibnamefont  [1]{#1}%
\providecommand \bibfnamefont [1]{#1}%
\providecommand \citenamefont [1]{#1}%
\providecommand \href@noop [0]{\@secondoftwo}%
\providecommand \href [0]{\begingroup \@sanitize@url \@href}%
\providecommand \@href[1]{\@@startlink{#1}\@@href}%
\providecommand \@@href[1]{\endgroup#1\@@endlink}%
\providecommand \@sanitize@url [0]{\catcode `\\12\catcode `\$12\catcode
  `\&12\catcode `\#12\catcode `\^12\catcode `\_12\catcode `\%12\relax}%
\providecommand \@@startlink[1]{}%
\providecommand \@@endlink[0]{}%
\providecommand \url  [0]{\begingroup\@sanitize@url \@url }%
\providecommand \@url [1]{\endgroup\@href {#1}{\urlprefix }}%
\providecommand \urlprefix  [0]{URL }%
\providecommand \Eprint [0]{\href }%
\providecommand \doibase [0]{http://dx.doi.org/}%
\providecommand \selectlanguage [0]{\@gobble}%
\providecommand \bibinfo  [0]{\@secondoftwo}%
\providecommand \bibfield  [0]{\@secondoftwo}%
\providecommand \translation [1]{[#1]}%
\providecommand \BibitemOpen [0]{}%
\providecommand \bibitemStop [0]{}%
\providecommand \bibitemNoStop [0]{.\EOS\space}%
\providecommand \EOS [0]{\spacefactor3000\relax}%
\providecommand \BibitemShut  [1]{\csname bibitem#1\endcsname}%
\let\auto@bib@innerbib\@empty
\bibitem [{\citenamefont {Schmertmann}(1991)}]{Schmertmann1991}%
  \BibitemOpen
  \bibfield  {author} {\bibinfo {author} {\bibfnamefont {J.~H.}\ \bibnamefont
  {Schmertmann}},\ }\href@noop {} {\bibfield  {journal} {\bibinfo  {journal}
  {J. Geotech. Eng.}\ }\textbf {\bibinfo {volume} {117}},\ \bibinfo {pages}
  {1288} (\bibinfo {year} {1991})}\BibitemShut {NoStop}%
\bibitem [{\citenamefont {Vandamme}\ and\ \citenamefont
  {Ulm}(2009)}]{Vandamme2009}%
  \BibitemOpen
  \bibfield  {author} {\bibinfo {author} {\bibfnamefont {M.}~\bibnamefont
  {Vandamme}}\ and\ \bibinfo {author} {\bibfnamefont {F.-J.}\ \bibnamefont
  {Ulm}},\ }\href@noop {} {\bibfield  {journal} {\bibinfo  {journal} {Proc.
  Natl. Acad. Sci. U. S. A.}\ }\textbf {\bibinfo {volume} {106}},\ \bibinfo
  {pages} {10552} (\bibinfo {year} {2009})}\BibitemShut {NoStop}%
\bibitem [{\citenamefont {O'Hern}\ \emph {et~al.}(2003)\citenamefont {O'Hern},
  \citenamefont {Silbert},\ and\ \citenamefont {Nagel}}]{OHern2003}%
  \BibitemOpen
  \bibfield  {author} {\bibinfo {author} {\bibfnamefont {C.~S.}\ \bibnamefont
  {O'Hern}}, \bibinfo {author} {\bibfnamefont {L.~E.}\ \bibnamefont {Silbert}},
  \ and\ \bibinfo {author} {\bibfnamefont {S.~R.}\ \bibnamefont {Nagel}},\
  }\href@noop {} {\bibfield  {journal} {\bibinfo  {journal} {Phys. Rev. E}\
  }\textbf {\bibinfo {volume} {68}},\ \bibinfo {pages} {011306} (\bibinfo
  {year} {2003})}\BibitemShut {NoStop}%
\bibitem [{\citenamefont {Majmudar}\ \emph {et~al.}(2007)\citenamefont
  {Majmudar}, \citenamefont {Sperl}, \citenamefont {Luding},\ and\
  \citenamefont {Behringer}}]{Majmudar2007}%
  \BibitemOpen
  \bibfield  {author} {\bibinfo {author} {\bibfnamefont {T.~S.}\ \bibnamefont
  {Majmudar}}, \bibinfo {author} {\bibfnamefont {M.}~\bibnamefont {Sperl}},
  \bibinfo {author} {\bibfnamefont {S.}~\bibnamefont {Luding}}, \ and\ \bibinfo
  {author} {\bibfnamefont {R.~P.}\ \bibnamefont {Behringer}},\ }\href@noop {}
  {\bibfield  {journal} {\bibinfo  {journal} {Phys. Rev. Lett.}\ }\textbf
  {\bibinfo {volume} {98}},\ \bibinfo {pages} {1} (\bibinfo {year}
  {2007})}\BibitemShut {NoStop}%
\bibitem [{\citenamefont {Bi}\ \emph {et~al.}(2011)\citenamefont {Bi},
  \citenamefont {Zhang}, \citenamefont {Chakraborty},\ and\ \citenamefont
  {Behringer}}]{Bi2011}%
  \BibitemOpen
  \bibfield  {author} {\bibinfo {author} {\bibfnamefont {D.}~\bibnamefont
  {Bi}}, \bibinfo {author} {\bibfnamefont {J.}~\bibnamefont {Zhang}}, \bibinfo
  {author} {\bibfnamefont {B.}~\bibnamefont {Chakraborty}}, \ and\ \bibinfo
  {author} {\bibfnamefont {R.~P.}\ \bibnamefont {Behringer}},\ }\href@noop {}
  {\bibfield  {journal} {\bibinfo  {journal} {Nature}\ }\textbf {\bibinfo
  {volume} {480}},\ \bibinfo {pages} {355} (\bibinfo {year}
  {2011})}\BibitemShut {NoStop}%
\bibitem [{\citenamefont {Dagois-Bohy}\ \emph {et~al.}(2012)\citenamefont
  {Dagois-Bohy}, \citenamefont {Tighe}, \citenamefont {Simon}, \citenamefont
  {Henkes},\ and\ \citenamefont {van Hecke}}]{Dagois-Bohy2012}%
  \BibitemOpen
  \bibfield  {author} {\bibinfo {author} {\bibfnamefont {S.}~\bibnamefont
  {Dagois-Bohy}}, \bibinfo {author} {\bibfnamefont {B.~P.}\ \bibnamefont
  {Tighe}}, \bibinfo {author} {\bibfnamefont {J.}~\bibnamefont {Simon}},
  \bibinfo {author} {\bibfnamefont {S.}~\bibnamefont {Henkes}}, \ and\ \bibinfo
  {author} {\bibfnamefont {M.}~\bibnamefont {van Hecke}},\ }\href@noop {}
  {\bibfield  {journal} {\bibinfo  {journal} {Phys. Rev. Lett.}\ }\textbf
  {\bibinfo {volume} {109}},\ \bibinfo {pages} {095703} (\bibinfo {year}
  {2012})}\BibitemShut {NoStop}%
\bibitem [{\citenamefont {{Da Cruz}}\ \emph {et~al.}(2005)\citenamefont {{Da
  Cruz}}, \citenamefont {Emam}, \citenamefont {Prochnow}, \citenamefont
  {Roux},\ and\ \citenamefont {Chevoir}}]{DaCruz2005}%
  \BibitemOpen
  \bibfield  {author} {\bibinfo {author} {\bibfnamefont {F.}~\bibnamefont {{Da
  Cruz}}}, \bibinfo {author} {\bibfnamefont {S.}~\bibnamefont {Emam}}, \bibinfo
  {author} {\bibfnamefont {M.}~\bibnamefont {Prochnow}}, \bibinfo {author}
  {\bibfnamefont {J.~N.}\ \bibnamefont {Roux}}, \ and\ \bibinfo {author}
  {\bibfnamefont {F.}~\bibnamefont {Chevoir}},\ }\href@noop {} {\bibfield
  {journal} {\bibinfo  {journal} {Physical Review E - Statistical, Nonlinear,
  and Soft Matter Physics}\ }\textbf {\bibinfo {volume} {72}},\ \bibinfo
  {pages} {1} (\bibinfo {year} {2005})}\BibitemShut {NoStop}%
\bibitem [{\citenamefont {Nedderman}(1992)}]{Nedderman:1992aa}%
  \BibitemOpen
  \bibfield  {author} {\bibinfo {author} {\bibfnamefont {R.~M.}\ \bibnamefont
  {Nedderman}},\ }\href@noop {} {\emph {\bibinfo {title} {{Statics and
  kinematics of granular materials}}}}\ (\bibinfo  {publisher} {Cambridge
  University Press},\ \bibinfo {year} {1992})\BibitemShut {NoStop}%
\bibitem [{\citenamefont {Jop}\ \emph {et~al.}(2006)\citenamefont {Jop},
  \citenamefont {Forterre},\ and\ \citenamefont {Pouliquen}}]{Jop2006}%
  \BibitemOpen
  \bibfield  {author} {\bibinfo {author} {\bibfnamefont {P.}~\bibnamefont
  {Jop}}, \bibinfo {author} {\bibfnamefont {Y.}~\bibnamefont {Forterre}}, \
  and\ \bibinfo {author} {\bibfnamefont {O.}~\bibnamefont {Pouliquen}},\
  }\href@noop {} {\bibfield  {journal} {\bibinfo  {journal} {Nature}\ }\textbf
  {\bibinfo {volume} {441}},\ \bibinfo {pages} {727} (\bibinfo {year}
  {2006})}\BibitemShut {NoStop}%
\bibitem [{\citenamefont {Balmforth}\ \emph {et~al.}(2014)\citenamefont
  {Balmforth}, \citenamefont {Frigaard},\ and\ \citenamefont
  {Ovarlez}}]{Balmforth2014}%
  \BibitemOpen
  \bibfield  {author} {\bibinfo {author} {\bibfnamefont {N.~J.}\ \bibnamefont
  {Balmforth}}, \bibinfo {author} {\bibfnamefont {I.~A.}\ \bibnamefont
  {Frigaard}}, \ and\ \bibinfo {author} {\bibfnamefont {G.}~\bibnamefont
  {Ovarlez}},\ }\href
  {http://www.annualreviews.org/doi/abs/10.1146/annurev-fluid-010313-141424}
  {\bibfield  {journal} {\bibinfo  {journal} {Annu. Rev. Fluid Mech.}\ }\textbf
  {\bibinfo {volume} {46}},\ \bibinfo {pages} {121} (\bibinfo {year}
  {2014})}\BibitemShut {NoStop}%
\bibitem [{\citenamefont {Coussot}\ \emph {et~al.}(2002)\citenamefont
  {Coussot}, \citenamefont {Nguyen}, \citenamefont {Huynh},\ and\ \citenamefont
  {Bonn}}]{Coussot2002}%
  \BibitemOpen
  \bibfield  {author} {\bibinfo {author} {\bibfnamefont {P.}~\bibnamefont
  {Coussot}}, \bibinfo {author} {\bibfnamefont {Q.~D.}\ \bibnamefont {Nguyen}},
  \bibinfo {author} {\bibfnamefont {H.~T.}\ \bibnamefont {Huynh}}, \ and\
  \bibinfo {author} {\bibfnamefont {D.}~\bibnamefont {Bonn}},\ }\href@noop {}
  {\bibfield  {journal} {\bibinfo  {journal} {Phys. Rev. Lett.}\ }\textbf
  {\bibinfo {volume} {88}},\ \bibinfo {pages} {175501} (\bibinfo {year}
  {2002})}\BibitemShut {NoStop}%
\bibitem [{\citenamefont {Bursac}\ \emph {et~al.}(2005)\citenamefont {Bursac},
  \citenamefont {Lenormand}, \citenamefont {Fabry}, \citenamefont {Oliver},
  \citenamefont {Weitz}, \citenamefont {Viasnoff}, \citenamefont {Butler},\
  and\ \citenamefont {Fredberg}}]{Bursac2005}%
  \BibitemOpen
  \bibfield  {author} {\bibinfo {author} {\bibfnamefont {P.}~\bibnamefont
  {Bursac}}, \bibinfo {author} {\bibfnamefont {G.}~\bibnamefont {Lenormand}},
  \bibinfo {author} {\bibfnamefont {B.}~\bibnamefont {Fabry}}, \bibinfo
  {author} {\bibfnamefont {M.}~\bibnamefont {Oliver}}, \bibinfo {author}
  {\bibfnamefont {D.~A.}\ \bibnamefont {Weitz}}, \bibinfo {author}
  {\bibfnamefont {V.}~\bibnamefont {Viasnoff}}, \bibinfo {author}
  {\bibfnamefont {J.~P.}\ \bibnamefont {Butler}}, \ and\ \bibinfo {author}
  {\bibfnamefont {J.~J.}\ \bibnamefont {Fredberg}},\ }\href@noop {} {\bibfield
  {journal} {\bibinfo  {journal} {Nat. Mater.}\ }\textbf {\bibinfo {volume}
  {4}},\ \bibinfo {pages} {557} (\bibinfo {year} {2005})}\BibitemShut {NoStop}%
\bibitem [{\citenamefont {Houssais}\ \emph {et~al.}(2015)\citenamefont
  {Houssais}, \citenamefont {Ortiz}, \citenamefont {Durian},\ and\
  \citenamefont {Jerolmack}}]{Houssais2015}%
  \BibitemOpen
  \bibfield  {author} {\bibinfo {author} {\bibfnamefont {M.}~\bibnamefont
  {Houssais}}, \bibinfo {author} {\bibfnamefont {C.~P.}\ \bibnamefont {Ortiz}},
  \bibinfo {author} {\bibfnamefont {D.~J.}\ \bibnamefont {Durian}}, \ and\
  \bibinfo {author} {\bibfnamefont {D.~J.}\ \bibnamefont {Jerolmack}},\
  }\href@noop {} {\bibfield  {journal} {\bibinfo  {journal} {Nat. Commun.}\
  }\textbf {\bibinfo {volume} {6}},\ \bibinfo {pages} {6527} (\bibinfo {year}
  {2015})}\BibitemShut {NoStop}%
\bibitem [{\citenamefont {Nguyen}\ \emph {et~al.}(2011)\citenamefont {Nguyen},
  \citenamefont {Darnige}, \citenamefont {Bruand},\ and\ \citenamefont
  {Clement}}]{Nguyen2011}%
  \BibitemOpen
  \bibfield  {author} {\bibinfo {author} {\bibfnamefont {V.~B.}\ \bibnamefont
  {Nguyen}}, \bibinfo {author} {\bibfnamefont {T.}~\bibnamefont {Darnige}},
  \bibinfo {author} {\bibfnamefont {A.}~\bibnamefont {Bruand}}, \ and\ \bibinfo
  {author} {\bibfnamefont {E.}~\bibnamefont {Clement}},\ }\href@noop {}
  {\bibfield  {journal} {\bibinfo  {journal} {Phys. Rev. Lett.}\ }\textbf
  {\bibinfo {volume} {107}},\ \bibinfo {pages} {1} (\bibinfo {year}
  {2011})}\BibitemShut {NoStop}%
\bibitem [{\citenamefont {Amon}\ \emph {et~al.}(2012)\citenamefont {Amon},
  \citenamefont {Nguyen}, \citenamefont {Bruand}, \citenamefont {Crassous},\
  and\ \citenamefont {Cl{\'{e}}ment}}]{Amon2012}%
  \BibitemOpen
  \bibfield  {author} {\bibinfo {author} {\bibfnamefont {A.}~\bibnamefont
  {Amon}}, \bibinfo {author} {\bibfnamefont {V.~B.}\ \bibnamefont {Nguyen}},
  \bibinfo {author} {\bibfnamefont {A.}~\bibnamefont {Bruand}}, \bibinfo
  {author} {\bibfnamefont {J.}~\bibnamefont {Crassous}}, \ and\ \bibinfo
  {author} {\bibfnamefont {E.}~\bibnamefont {Cl{\'{e}}ment}},\ }\href@noop {}
  {\bibfield  {journal} {\bibinfo  {journal} {Phys. Rev. Lett.}\ }\textbf
  {\bibinfo {volume} {108}},\ \bibinfo {pages} {1} (\bibinfo {year}
  {2012})}\BibitemShut {NoStop}%
\bibitem [{\citenamefont {Hartley}\ and\ \citenamefont
  {Behringer}(2003)}]{Hartley2003}%
  \BibitemOpen
  \bibfield  {author} {\bibinfo {author} {\bibfnamefont {R.~R.}\ \bibnamefont
  {Hartley}}\ and\ \bibinfo {author} {\bibfnamefont {R.~P.}\ \bibnamefont
  {Behringer}},\ }\href@noop {} {\bibfield  {journal} {\bibinfo  {journal}
  {Nature}\ }\textbf {\bibinfo {volume} {421}},\ \bibinfo {pages} {928}
  (\bibinfo {year} {2003})}\BibitemShut {NoStop}%
\bibitem [{\citenamefont {Nasuno}\ \emph {et~al.}(1998)\citenamefont {Nasuno},
  \citenamefont {Kudrolli}, \citenamefont {Bak},\ and\ \citenamefont
  {Gollub}}]{Nasuno1998}%
  \BibitemOpen
  \bibfield  {author} {\bibinfo {author} {\bibfnamefont {S.}~\bibnamefont
  {Nasuno}}, \bibinfo {author} {\bibfnamefont {A.}~\bibnamefont {Kudrolli}},
  \bibinfo {author} {\bibfnamefont {A.}~\bibnamefont {Bak}}, \ and\ \bibinfo
  {author} {\bibfnamefont {J.~P.}\ \bibnamefont {Gollub}},\ }\href@noop {}
  {\bibfield  {journal} {\bibinfo  {journal} {Phys. Rev. E}\ }\textbf {\bibinfo
  {volume} {58}},\ \bibinfo {pages} {2161} (\bibinfo {year}
  {1998})}\BibitemShut {NoStop}%
\bibitem [{\citenamefont {Bruji{\'{c}}}\ \emph {et~al.}(2005)\citenamefont
  {Bruji{\'{c}}}, \citenamefont {Wang}, \citenamefont {Song}, \citenamefont
  {Johnson}, \citenamefont {Sindt},\ and\ \citenamefont {Makse}}]{Brujic2005}%
  \BibitemOpen
  \bibfield  {author} {\bibinfo {author} {\bibfnamefont {J.}~\bibnamefont
  {Bruji{\'{c}}}}, \bibinfo {author} {\bibfnamefont {P.}~\bibnamefont {Wang}},
  \bibinfo {author} {\bibfnamefont {C.}~\bibnamefont {Song}}, \bibinfo {author}
  {\bibfnamefont {D.~L.}\ \bibnamefont {Johnson}}, \bibinfo {author}
  {\bibfnamefont {O.}~\bibnamefont {Sindt}}, \ and\ \bibinfo {author}
  {\bibfnamefont {H.~A.}\ \bibnamefont {Makse}},\ }\href@noop {} {\bibfield
  {journal} {\bibinfo  {journal} {Phys. Rev. Lett.}\ }\textbf {\bibinfo
  {volume} {95}},\ \bibinfo {pages} {1} (\bibinfo {year} {2005})}\BibitemShut
  {NoStop}%
\bibitem [{\citenamefont {Richard}\ \emph {et~al.}(2005)\citenamefont
  {Richard}, \citenamefont {Nicodemi}, \citenamefont {Delannay}, \citenamefont
  {Ribi{\`{e}}re},\ and\ \citenamefont {Bideau}}]{Richard2005}%
  \BibitemOpen
  \bibfield  {author} {\bibinfo {author} {\bibfnamefont {P.}~\bibnamefont
  {Richard}}, \bibinfo {author} {\bibfnamefont {M.}~\bibnamefont {Nicodemi}},
  \bibinfo {author} {\bibfnamefont {R.}~\bibnamefont {Delannay}}, \bibinfo
  {author} {\bibfnamefont {P.}~\bibnamefont {Ribi{\`{e}}re}}, \ and\ \bibinfo
  {author} {\bibfnamefont {D.}~\bibnamefont {Bideau}},\ }\href@noop {}
  {\bibfield  {journal} {\bibinfo  {journal} {Nat. Mater.}\ }\textbf {\bibinfo
  {volume} {4}},\ \bibinfo {pages} {121} (\bibinfo {year} {2005})}\BibitemShut
  {NoStop}%
\bibitem [{\citenamefont {Knight}\ \emph {et~al.}(1995)\citenamefont {Knight},
  \citenamefont {Fandrich}, \citenamefont {Lau}, \citenamefont {Jaeger},\ and\
  \citenamefont {Nagel}}]{Knight1995}%
  \BibitemOpen
  \bibfield  {author} {\bibinfo {author} {\bibfnamefont {J.~B.}\ \bibnamefont
  {Knight}}, \bibinfo {author} {\bibfnamefont {C.~G.}\ \bibnamefont
  {Fandrich}}, \bibinfo {author} {\bibfnamefont {C.~N.}\ \bibnamefont {Lau}},
  \bibinfo {author} {\bibfnamefont {H.~M.}\ \bibnamefont {Jaeger}}, \ and\
  \bibinfo {author} {\bibfnamefont {S.~R.}\ \bibnamefont {Nagel}},\ }\href@noop
  {} {\bibfield  {journal} {\bibinfo  {journal} {Phys. Rev. E}\ }\textbf
  {\bibinfo {volume} {51}},\ \bibinfo {pages} {3957} (\bibinfo {year}
  {1995})}\BibitemShut {NoStop}%
\bibitem [{\citenamefont {Maloney}\ and\ \citenamefont
  {Lema{\^{\i}}tre}(2006)}]{Maloney2006}%
  \BibitemOpen
  \bibfield  {author} {\bibinfo {author} {\bibfnamefont {C.}~\bibnamefont
  {Maloney}}\ and\ \bibinfo {author} {\bibfnamefont {A.}~\bibnamefont
  {Lema{\^{\i}}tre}},\ }\href@noop {} {\bibfield  {journal} {\bibinfo
  {journal} {Phys. Rev. E}\ }\textbf {\bibinfo {volume} {74}},\ \bibinfo
  {pages} {016118} (\bibinfo {year} {2006})}\BibitemShut {NoStop}%
\bibitem [{\citenamefont {Schall}\ \emph {et~al.}(2007)\citenamefont {Schall},
  \citenamefont {Weitz},\ and\ \citenamefont {Spaepen}}]{Schall2007}%
  \BibitemOpen
  \bibfield  {author} {\bibinfo {author} {\bibfnamefont {P.}~\bibnamefont
  {Schall}}, \bibinfo {author} {\bibfnamefont {D.~A.}\ \bibnamefont {Weitz}}, \
  and\ \bibinfo {author} {\bibfnamefont {F.}~\bibnamefont {Spaepen}},\
  }\href@noop {} {\bibfield  {journal} {\bibinfo  {journal} {Science}\ }\textbf
  {\bibinfo {volume} {318}},\ \bibinfo {pages} {1895} (\bibinfo {year}
  {2007})}\BibitemShut {NoStop}%
\bibitem [{\citenamefont {Kamrin}\ and\ \citenamefont
  {Koval}(2012)}]{Kamrin2012}%
  \BibitemOpen
  \bibfield  {author} {\bibinfo {author} {\bibfnamefont {K.}~\bibnamefont
  {Kamrin}}\ and\ \bibinfo {author} {\bibfnamefont {G.}~\bibnamefont {Koval}},\
  }\href@noop {} {\bibfield  {journal} {\bibinfo  {journal} {Phys. Rev. Lett.}\
  }\textbf {\bibinfo {volume} {108}},\ \bibinfo {pages} {1} (\bibinfo {year}
  {2012})}\BibitemShut {NoStop}%
\bibitem [{\citenamefont {Bocquet}\ \emph {et~al.}(2009)\citenamefont
  {Bocquet}, \citenamefont {Colin},\ and\ \citenamefont
  {Ajdari}}]{Bocquet2009}%
  \BibitemOpen
  \bibfield  {author} {\bibinfo {author} {\bibfnamefont {L.}~\bibnamefont
  {Bocquet}}, \bibinfo {author} {\bibfnamefont {A.}~\bibnamefont {Colin}}, \
  and\ \bibinfo {author} {\bibfnamefont {A.}~\bibnamefont {Ajdari}},\
  }\href@noop {} {\bibfield  {journal} {\bibinfo  {journal} {Phys. Rev. Lett.}\
  }\textbf {\bibinfo {volume} {103}},\ \bibinfo {pages} {1} (\bibinfo {year}
  {2009})}\BibitemShut {NoStop}%
\bibitem [{\citenamefont {Kuhn}\ and\ \citenamefont
  {Mitchell}(1993)}]{Kuhn1993}%
  \BibitemOpen
  \bibfield  {author} {\bibinfo {author} {\bibfnamefont {M.~R.}\ \bibnamefont
  {Kuhn}}\ and\ \bibinfo {author} {\bibfnamefont {J.~K.}\ \bibnamefont
  {Mitchell}},\ }\href@noop {} {\bibfield  {journal} {\bibinfo  {journal} {J.
  Geotech. Eng.}\ }\textbf {\bibinfo {volume} {119}},\ \bibinfo {pages} {507}
  (\bibinfo {year} {1993})}\BibitemShut {NoStop}%
\bibitem [{\citenamefont {Smith}\ \emph {et~al.}(2014)\citenamefont {Smith},
  \citenamefont {Srivastava}, \citenamefont {Fisher},\ and\ \citenamefont
  {Alam}}]{Smith2014}%
  \BibitemOpen
  \bibfield  {author} {\bibinfo {author} {\bibfnamefont {K.~C.}\ \bibnamefont
  {Smith}}, \bibinfo {author} {\bibfnamefont {I.}~\bibnamefont {Srivastava}},
  \bibinfo {author} {\bibfnamefont {T.~S.}\ \bibnamefont {Fisher}}, \ and\
  \bibinfo {author} {\bibfnamefont {M.}~\bibnamefont {Alam}},\ }\href {\doibase
  10.1103/PhysRevE.89.042203} {\bibfield  {journal} {\bibinfo  {journal} {Phys.
  Rev. E}\ }\textbf {\bibinfo {volume} {89}},\ \bibinfo {pages} {042203}
  (\bibinfo {year} {2014})}\BibitemShut {NoStop}%
\bibitem [{\citenamefont {Smith}\ \emph {et~al.}(2010)\citenamefont {Smith},
  \citenamefont {Alam},\ and\ \citenamefont {Fisher}}]{Smith2010}%
  \BibitemOpen
  \bibfield  {author} {\bibinfo {author} {\bibfnamefont {K.~C.}\ \bibnamefont
  {Smith}}, \bibinfo {author} {\bibfnamefont {M.}~\bibnamefont {Alam}}, \ and\
  \bibinfo {author} {\bibfnamefont {T.~S.}\ \bibnamefont {Fisher}},\ }\href
  {\doibase 10.1103/PhysRevE.82.051304} {\bibfield  {journal} {\bibinfo
  {journal} {Phys. Rev. E}\ }\textbf {\bibinfo {volume} {82}},\ \bibinfo
  {pages} {1} (\bibinfo {year} {2010})}\BibitemShut {NoStop}%
\bibitem [{\citenamefont {V{\aa}gberg}\ \emph {et~al.}(2014)\citenamefont
  {V{\aa}gberg}, \citenamefont {Olsson},\ and\ \citenamefont
  {Teitel}}]{Vagberg2014}%
  \BibitemOpen
  \bibfield  {author} {\bibinfo {author} {\bibfnamefont {D.}~\bibnamefont
  {V{\aa}gberg}}, \bibinfo {author} {\bibfnamefont {P.}~\bibnamefont {Olsson}},
  \ and\ \bibinfo {author} {\bibfnamefont {S.}~\bibnamefont {Teitel}},\ }\href
  {\doibase 10.1103/PhysRevLett.113.148002} {\bibfield  {journal} {\bibinfo
  {journal} {Phys. Rev. Lett.}\ }\textbf {\bibinfo {volume} {113}},\ \bibinfo
  {pages} {148002} (\bibinfo {year} {2014})}\BibitemShut {NoStop}%
\bibitem [{Note1()}]{Note1}%
  \BibitemOpen
  \bibinfo {note} {See Supplemental Material for detailed derivation and
  dimensional analysis of the equations of motion.}\BibitemShut {Stop}%
\bibitem [{\citenamefont {Souza}\ and\ \citenamefont
  {Martins}(1997)}]{Souza1997}%
  \BibitemOpen
  \bibfield  {author} {\bibinfo {author} {\bibfnamefont {I.}~\bibnamefont
  {Souza}}\ and\ \bibinfo {author} {\bibfnamefont {J.}~\bibnamefont
  {Martins}},\ }\href {\doibase 10.1103/PhysRevB.55.8733} {\bibfield  {journal}
  {\bibinfo  {journal} {Phy. Rev. B}\ }\textbf {\bibinfo {volume} {55}},\
  \bibinfo {pages} {8733} (\bibinfo {year} {1997})}\BibitemShut {NoStop}%
\bibitem [{\citenamefont {Findley}\ and\ \citenamefont
  {Lai}(1976)}]{Findley:1976aa}%
  \BibitemOpen
  \bibfield  {author} {\bibinfo {author} {\bibfnamefont {W.~N.}\ \bibnamefont
  {Findley}}\ and\ \bibinfo {author} {\bibfnamefont {J.~S.}\ \bibnamefont
  {Lai}},\ }\href@noop {} {\emph {\bibinfo {title} {Creep and relaxation of
  nonlinear viscoelastic materials}}}\ (\bibinfo  {publisher} {Dover
  Publications},\ \bibinfo {address} {New York},\ \bibinfo {year}
  {1976})\BibitemShut {NoStop}%
\bibitem [{\citenamefont {Siebenb{\"{u}}rger}\ \emph
  {et~al.}(2012)\citenamefont {Siebenb{\"{u}}rger}, \citenamefont {Ballauff},\
  and\ \citenamefont {Voigtmann}}]{Siebenburger2012}%
  \BibitemOpen
  \bibfield  {author} {\bibinfo {author} {\bibfnamefont {M.}~\bibnamefont
  {Siebenb{\"{u}}rger}}, \bibinfo {author} {\bibfnamefont {M.}~\bibnamefont
  {Ballauff}}, \ and\ \bibinfo {author} {\bibfnamefont {T.}~\bibnamefont
  {Voigtmann}},\ }\href {\doibase 10.1103/PhysRevLett.108.255701} {\bibfield
  {journal} {\bibinfo  {journal} {Phys. Rev. Lett.}\ }\textbf {\bibinfo
  {volume} {108}},\ \bibinfo {pages} {1} (\bibinfo {year} {2012})}\BibitemShut
  {NoStop}%
\bibitem [{\citenamefont {Bowman}\ and\ \citenamefont
  {Soga}(2005)}]{Bowman2005}%
  \BibitemOpen
  \bibfield  {author} {\bibinfo {author} {\bibfnamefont {E.~T.}\ \bibnamefont
  {Bowman}}\ and\ \bibinfo {author} {\bibfnamefont {K.}~\bibnamefont {Soga}},\
  }\href {\doibase 10.1139/t05-063} {\bibfield  {journal} {\bibinfo  {journal}
  {Can. Geotech. J.}\ }\textbf {\bibinfo {volume} {42}},\ \bibinfo {pages}
  {1391} (\bibinfo {year} {2005})}\BibitemShut {NoStop}%
\bibitem [{\citenamefont {Falk}\ and\ \citenamefont {Langer}(1998)}]{Falk1998}%
  \BibitemOpen
  \bibfield  {author} {\bibinfo {author} {\bibfnamefont {M.}~\bibnamefont
  {Falk}}\ and\ \bibinfo {author} {\bibfnamefont {J.}~\bibnamefont {Langer}},\
  }\href {\doibase 10.1103/PhysRevE.57.7192} {\bibfield  {journal} {\bibinfo
  {journal} {Phys. Rev. E}\ }\textbf {\bibinfo {volume} {57}},\ \bibinfo
  {pages} {7192} (\bibinfo {year} {1998})}\BibitemShut {NoStop}%
\bibitem [{\citenamefont {Wang}\ \emph {et~al.}(2013)\citenamefont {Wang},
  \citenamefont {Mao}, \citenamefont {Shan}, \citenamefont {Dao}, \citenamefont
  {Li}, \citenamefont {Sun}, \citenamefont {Ma},\ and\ \citenamefont
  {Suresh}}]{Wang2013}%
  \BibitemOpen
  \bibfield  {author} {\bibinfo {author} {\bibfnamefont {C.-C.}\ \bibnamefont
  {Wang}}, \bibinfo {author} {\bibfnamefont {Y.-W.}\ \bibnamefont {Mao}},
  \bibinfo {author} {\bibfnamefont {Z.-W.}\ \bibnamefont {Shan}}, \bibinfo
  {author} {\bibfnamefont {M.}~\bibnamefont {Dao}}, \bibinfo {author}
  {\bibfnamefont {J.}~\bibnamefont {Li}}, \bibinfo {author} {\bibfnamefont
  {J.}~\bibnamefont {Sun}}, \bibinfo {author} {\bibfnamefont {E.}~\bibnamefont
  {Ma}}, \ and\ \bibinfo {author} {\bibfnamefont {S.}~\bibnamefont {Suresh}},\
  }\href {\doibase 10.1073/pnas.1320235110} {\bibfield  {journal} {\bibinfo
  {journal} {Proc. Natl. Acad. Sci. U. S. A.}\ }\textbf {\bibinfo {volume}
  {110}},\ \bibinfo {pages} {19725} (\bibinfo {year} {2013})}\BibitemShut
  {NoStop}%
\bibitem [{\citenamefont {Jensen}\ \emph {et~al.}(2014)\citenamefont {Jensen},
  \citenamefont {Weitz},\ and\ \citenamefont {Spaepen}}]{Jensen2014}%
  \BibitemOpen
  \bibfield  {author} {\bibinfo {author} {\bibfnamefont {K.~E.}\ \bibnamefont
  {Jensen}}, \bibinfo {author} {\bibfnamefont {D.~A.}\ \bibnamefont {Weitz}}, \
  and\ \bibinfo {author} {\bibfnamefont {F.}~\bibnamefont {Spaepen}},\ }\href
  {\doibase 10.1103/PhysRevE.90.042305} {\bibfield  {journal} {\bibinfo
  {journal} {Phys. Rev. E}\ }\textbf {\bibinfo {volume} {90}},\ \bibinfo
  {pages} {042305} (\bibinfo {year} {2014})}\BibitemShut {NoStop}%
\bibitem [{\citenamefont {Chattoraj}\ and\ \citenamefont
  {Lema{\^{\i}}tre}(2013)}]{Chattoraj2013}%
  \BibitemOpen
  \bibfield  {author} {\bibinfo {author} {\bibfnamefont {J.}~\bibnamefont
  {Chattoraj}}\ and\ \bibinfo {author} {\bibfnamefont {A.}~\bibnamefont
  {Lema{\^{\i}}tre}},\ }\href {\doibase 10.1103/PhysRevLett.111.066001}
  {\bibfield  {journal} {\bibinfo  {journal} {Phys. Rev. Lett.}\ }\textbf
  {\bibinfo {volume} {111}},\ \bibinfo {pages} {1} (\bibinfo {year}
  {2013})}\BibitemShut {NoStop}%
\end{thebibliography}%


\begin{thebibliography}{3}
\expandafter\ifx\csname natexlab\endcsname\relax\def\natexlab#1{#1}\fi
\expandafter\ifx\csname bibnamefont\endcsname\relax
  \def\bibnamefont#1{#1}\fi
\expandafter\ifx\csname bibfnamefont\endcsname\relax
  \def\bibfnamefont#1{#1}\fi
\expandafter\ifx\csname citenamefont\endcsname\relax
  \def\citenamefont#1{#1}\fi
\expandafter\ifx\csname url\endcsname\relax
  \def\url#1{\texttt{#1}}\fi
\expandafter\ifx\csname urlprefix\endcsname\relax\def\urlprefix{URL }\fi
\providecommand{\bibinfo}[2]{#2}
\providecommand{\eprint}[2][]{\url{#2}}

\bibitem[{\citenamefont{Souza and Martins}(1997)}]{Souza1997}
\bibinfo{author}{\bibfnamefont{I.}~\bibnamefont{Souza}} \bibnamefont{and}
  \bibinfo{author}{\bibfnamefont{J.}~\bibnamefont{Martins}},
  \bibinfo{journal}{Phy. Rev. B} \textbf{\bibinfo{volume}{55}},
  \bibinfo{pages}{8733} (\bibinfo{year}{1997}).

\bibitem[{\citenamefont{Smith et~al.}(2014)\citenamefont{Smith, Srivastava,
  Fisher, and Alam}}]{Smith2014}
\bibinfo{author}{\bibfnamefont{K.~C.} \bibnamefont{Smith}},
  \bibinfo{author}{\bibfnamefont{I.}~\bibnamefont{Srivastava}},
  \bibinfo{author}{\bibfnamefont{T.~S.} \bibnamefont{Fisher}},
  \bibnamefont{and} \bibinfo{author}{\bibfnamefont{M.}~\bibnamefont{Alam}},
  \bibinfo{journal}{Phys. Rev. E} \textbf{\bibinfo{volume}{89}},
  \bibinfo{pages}{042203} (\bibinfo{year}{2014}).

\bibitem[{\citenamefont{Goldstein}(1980)}]{Goldstein}
\bibinfo{author}{\bibfnamefont{H.}~\bibnamefont{Goldstein}},
  \emph{\bibinfo{title}{Classical Mechanics}}
  (\bibinfo{publisher}{Addison-Wesley}, \bibinfo{address}{Reading, MA},
  \bibinfo{year}{1980}), \bibinfo{edition}{2nd} ed.

\end{thebibliography}

\end{document}


\title{Supplemental Information for the Letter {\em Local Plasticity as the Source of Creep and Slow Dynamics in Granular Materials.}}
\author{Ishan Srivastava}
\author{Timothy S Fisher}
\affiliation{School of Mechanical Engineering and Birck Nanotechnology Center, Purdue University, West Lafayette, IN 47907, USA}

\maketitle

\section{Equations of Motion}
In the main text, we defined the equations of motion of a granular system that is jammed at an external pressure $P$ and is subjected to a vanishingly low state of stress $\bm{\sigma}_{\text{ext}}$. Here, we give details about the derivation of these equations from a Lagrangian formulation of the system dynamics. Because the periodic cell boundaries evolve dynamically as a response to the external stress, a metric tensor $\bm{g}=\bm{h}^T\bm{h}$ is introduced as a dynamical variable \cite{Souza1997}. Here, $\bm{h}=[\bm{a}_{1},\bm{a}_{2},\bm{a}_{3}]$, where $\bm{a}_{1}$ ,$\bm{a}_{2}$, and $\bm{a}_{3}$ are the Bravais lattice vectors that define the periodic cell. This metric tensor is symmetric with $6$ degrees of freedom, and the choice of this dynamical variable introduces rotational invariance in the system dynamics. Because an infinite number of Bravais cell vectors can describe a periodic system, we choose a simple and unique form of a symmetric cell matrix $\bm{h} = \bm{g}^{1/2}$ \cite{Smith2014}. The volume of the cell $\Omega$  is given by $\text{det}(\bm{h})$. The position $\bm{r}_{i}$ of a grain $i$ in the Cartesian space can be defined by its lattice coordinates in the metric space $\bm{s}_{i}$ as: $\bm{r}_{i} = \bm{h}\bm{s}_{i}$. The angular position of a grain $i$ is defined by the rotation vector $\bm{\omega}_{i}$. 

Based on the formulations of Souza and Martins \cite{Souza1997}, we construct a Lagrangian from which the temporal equations of motion of $6N+6$ generalized coordinates ($3N$ granular translation $[\bm{s}] = (\bm{s}_{1},\bm{s}_{2},...,\bm{s}_{N})$, $3N$ granular rotation $[\bm{\omega}] = (\bm{\omega}_{1},\bm{\omega}_{2},...,\bm{\omega}_{N})$ and $6$ periodic cell degrees of freedom) are obtained. The total kinetic energy of the system is:
\begin{equation}
\mathcal{T} = \frac{1}{2}\sum_{i=1}^{N} m_{i} \dot{\bm{s}_{i}}^T \bm{g} \dot{\bm{s}_{i}} + \frac{1}{2}\sum_{i=1}^{N} \dot{\bm{\omega}}_{i}^T \bm{I}_{i} \dot{\bm{\omega}}_{i} + \frac{1}{2}W_{c}\Omega^{2} \text{Tr}\left(\dot{\bm{g}} \bm{g}^{-1}\dot{\bm{g}}\bm{g}^{-1}\right).
\label{eq1}
\end{equation}
Here, $m_{i}$ is the mass and $\bm{I}_{i}$ is the inertia tensor of grain $i$. The third term in the equation is a fictitious kinetic energy of the periodic cell where $W_c$ is a fictitious cell mass with units of mass times length$^{-4}$ \cite{Souza1997}. `Tr' represents the trace of a matrix. The total potential energy of the system is:
\begin{equation}
\mathcal{V} = U_{\text{g}}([\bm{s}],[\bm{\omega}],\bm{g}) + P\Omega + \text{Tr}\left(\widetilde{\bm{\sigma}}_{\text{ext}}\bm{g}\right), 
\label{eq2}
\end{equation}
with contributions from elastic granular forces and moments $U_{\text{g}}([\bm{s}],[\bm{\omega}],\bm{g})$ and elastic work done by external stress on the periodic cell. The details of the calculation of the elastic forces and moments have been described previously in \cite{Smith2014}. The last two terms denote the work done by pressure $P$ and external stress $\bm{\sigma}_{\text{ext}}$. In Eq.~\ref{eq2}, $\widetilde{\bm{\sigma}}_{\text{ext}}$ is the external stress tensor in lattice coordinates \cite{Smith2014}. The Lagrangian of the system is thus defined as $\mathcal{L} = \mathcal{T} - \mathcal{V}$.

A Rayleigh dissipation function $\mathcal{D}$ \cite{Goldstein} that is employed to model the frictional dissipation forces described in the main text is defined as:
\begin{equation}
\mathcal{D} = \frac{1}{2}\sum_{i=1}^{N}D_{g}\left(\dot{\bm{s}_{i}}^T \dot{\bm{s}_{i}} + \dot{\bm{\omega}_{i}}^T \dot{\bm{\omega}_{i}}\right) + D_{c}\text{Tr}(\dot{\bm{g}}^{T}\dot{\bm{g}}),
\label{eq3}
\end{equation}
where $D_{g}$ and $D_{c}$ and the dissipation constants associated with the motion of grains and periodic cell respectively. Note that the dissipation is applied only to non-affine motion of the granular translation, as depicted in the first term of the dissipation function. Recall that $\bm{r}_{i}=\bm{h}\bm{s}_{i}$, and therefore the total velocity of a grain $i$ is $\dot{\bm{r}}_{i} = \bm{h}\dot{\bm{s}}_{i} + \dot{\bm{h}}\bm{s}_{i}$. Here the first term defines the fluctuating non-affine velocity of grain $i$ in addition to the velocity contribution of the affine homogenous system-level deformation defined in the second term. 
The Lagrangian equation of motion is given by:
\begin{equation}
\frac{d}{dt}\left(\frac{\partial \mathcal{L} }{\partial \dot{q}}\right) - \frac{\partial \mathcal{L}}{\partial q} + \frac{\partial \mathcal{D}}{\partial \dot{q}}=0,
\label{eq4}
\end{equation}
where $q$ is the set of generalized coordinates. Therefore, the equation of translational motion of a grain $i$ is:
\begin{equation}
m_{i}\bm{g}\ddot{\bm{s}}_{i} + m_{i}\dot{\bm{g}}\dot{\bm{s}}_{i} + D_{g}\dot{\bm{s}}_{i} = \widetilde{\mathbf{F}}_{i}^{e},
\label{eq5}
\end{equation}
where $\widetilde{\mathbf{F}}_{i}^{e} = -\frac{\partial U_{g}}{\partial \bm{s}_{i}}$ is the net elastic force on grain $i$ in reciprocal lattice coordinates \cite{Souza1997}. The equation of rotational motion of grain $i$ is given as:
\begin{equation}
\bm{I}_{i} \ddot{\bm{\omega}}_{i} + D_{g}\dot{\bm{\omega}}_{i} = \mathbf{M}_{i}^{e},
\label{eq6}
\end{equation}
where $\mathbf{M}_{i}^{e} = -\frac{\partial U_{g}}{\partial \bm{\omega}_{i}}$ is the net elastic moment on grain $i$. Since the periodic cell is also represented as a dynamical variable in the set of generalized coordinates, the Lagrangian equation for motion of the periodic cell is given as:
\begin{equation}
W_{c}\bm{g}^{-1}\ddot{\bm{g}}\bm{g}^{-1} + \frac{1}{\Omega^{2}}D_{c}\dot{\bm{g}} = \frac{1}{2\Omega^{2}}\left(\widetilde{\bm{\sigma}}_{\text{int}} - \widetilde{\bm{\sigma}}_{\text{app}}\right) + W_{c}\bm{g}^{-1}\left[\dot{\bm{g}}\bm{g}^{-1}\dot{\bm{g}} - \text{Tr}\left(\bm{g}^{-1}\dot{\bm{g}}\right)\dot{\bm{g}}\right]\bm{g}^{-1} + \frac{W_{c}}{2}\bm{g}^{-1}\text{Tr}\left(\dot{\bm{g}}\bm{g}^{-1}\dot{\bm{g}}\bm{g}^{-1}\right).
\label{eq7}
\end{equation}
Here, $\widetilde{\bm{\sigma}}_{\text{int}}=-\frac{\partial U_{g}}{\partial \bm{g}}$ is the internal stress tensor \cite{Smith2014} in lattice coordinates. The lattice coordinates of the external stress tensor $\bm{\sigma}_{\text{app}} = \bm{\sigma}_{\text{ext}} + P\mathbf{I}$ are expressed in the tensor $\widetilde{\bm{\sigma}}_{\text{app}}$. The transformation of a stress tensor from Cartesian $\bm{\sigma}$ to lattice coordinates $\widetilde{\bm{\sigma}}$ is expressed as \cite{Souza1997}:
\begin{equation}
\widetilde{\bm{\sigma}} = \left(\text{det}\bm{h}\right)\bm{h}^{-1}\bm{\sigma}\bm{h}^{-T},
\label{eq8}
\end{equation}

\section{Dimensional Analysis}

 A dimensionless contact stiffness number defined as $\kappa = P/Y$ determines the softness of the granular system, where $P$ is the external jamming (confining) pressure. The limit $\kappa \to 0$ represents the limit of hard grains, whereas $\kappa$ increases with increasing granular softness. The present simulations correspond to $\kappa < 10^{-4}$ and therefore, the system is assumed to contain hard grains. 

An inertial relaxation time scale $\tau_{I}^{g} = \sqrt{m/\left(\kappa Pa\right)}$ and a dissipative relaxation time scale $\tau_{D}^{g} = D_{g}/\left(\kappa Pa^{3}\right)$ for granular motion can be extracted from Eqs.~\ref{eq5} and~\ref{eq6} by considering that the energy of a grain-grain elastic contact scales as $U \sim \kappa P a^{3}$ for a harmonic overlap potential.  In the present simulation of overdamped granular dynamics, the limit $\tau_{I}^{g} \ll \tau_{D}^{g}$ is assumed, and the equations of granular motion reduce to:
\begin{equation}
D_{g}\dot{\bm{s}}_{i} = \widetilde{\mathbf{F}}_{i}^{e} \qquad\text{and}\qquad D_{g}\dot{\bm{\omega}}_{i} = \mathbf{M}_{i}^{e}.
\label{eq9}
\end{equation}
The global strain rate is calculated by measuring the deformation of periodic cell boundaries as described in Eq.~\ref{eq7}. In the limit of small applied external stress (below yield stress), two time scales corresponding to inertial relaxation time of the periodic cell $\tau_{I}^{c} = \sqrt{W_{c}a^{3}/P}$ and dissipative relaxation of the periodic cell $\tau_{D}^{c} = D_{c}a/P$ can be extracted. In the present simulations of overdamped dynamics, the limit $\tau_{I}^{c} \ll \tau_{D}^{c}$ is assumed, and the equation of cell motion reduces to:
\begin{equation}
D_{c}\dot{\bm{g}} + \left(\widetilde{\bm{\sigma}}_{\text{app}} - \widetilde{\bm{\sigma}}_{\text{int}}\right) = 0.
\label{eq10}
\end{equation}
Based on this analysis, the time in the main text is normalized by $\tau_{D}^{g}$, and the dissipation constant for periodic cell motion is prescribed as $D_{c} = D_{g}/\left(\kappa a^{4}\right)$.

\bibliography{supp}